Reproductive Tradeoffs In Uncertain Environments:
Explaining The Evolution Of Cultural Elaboration

Mark Madsen, Carl Lipo, and Michael Cannon

Mark Madsen
Emergent Media, Inc.
411 Fairview Ave.
Suite 200
Seattle, WA 98109
Email: madsen@emergentmedia.com

Carl Lipo
Department of Anthropology
Box 353100
University of Washington, Seattle
Seattle, WA 98195-3100
Email: clipo@u.washington.edu

Mike Cannon
Department of Anthropology
Box 353100
University of Washington, Seattle
Seattle, WA 98195-3100
Email: mcannon@u.washington.edu

# ABSTRACT

Dunnell's (1989) explanation for cultural elaboration proposed that elaboration could be a consequence of selection within uncertain environments. He developed the theory to the extent that it performed well explaining the distribution of elaboration at regional scales within eastern North America. More detailed studies require more detailed development of the theory, so that additional hypotheses and implications can be deduced. We draw upon theories of selection in fluctuating environments and portions of life history theory to propose a formal model of selection for "waste" in uncertain environments. We use agent-based simulation to explore the sufficiency of the theory for explaining the maintenance of high levels of elaboration within populations, and deduce novel implications for archaeological evidence usable by others to test the utility of Dunnell's model in specific localities.

"Theory provides the maps that turn an uncoordinated set of experiments or computer simulations into a cumulative exploration"

-- Booker, et al. (1989) *Artificial Intelligence* (40): 235-282

## INTRODUCTION

Cultural elaboration, particularly in its most spectacular expressions in burial ceremonialism and monumental architecture, is a natural source of curiosity for archaeologists whenever and wherever it occurs. Who would not be interested in why Hopewell burials contain immense quantities of obsidian, copper, and other difficult-to-acquire materials, or why Moche monumental architecture and burials flourished for a time on the coast of Peru and then vanished? Indeed, the earliest archaeological work in many regions of the world has focused on monumental architecture and burials. Consequently, explanations for cultural elaboration are many and contain a common thread traceable to $19^{th}$ century cultural evolution (Dunnell 1989; Madsen *in press*; Madsen, Kornbacher and Sterling *this volume*). Until recently, however, there has been little effort aimed at explaining elaboration as a natural consequence of broader principles of human cultural evolution. To us, the critical question remains: why would people expend such enormous amounts of energy on tasks and objects seemingly unrelated to their survival or to reproduction?

If we accept the proposition that human cultural behavior evolves by the same Darwinian principles as the rest of the organic world, we are forced to seek an explanation for elaboration using these same principles, namely natural selection, cultural transmission, and drift (Dunnell 1980, 1989; O'Brien and Holland 1990). Exploring the outlines of such an explanation is the purpose of the present chapter. To be clear, our analysis in the present context is not a complete explanation of cultural elaboration in all its forms and conceptions. A general *explanation* cannot exist because evolution is an endless interplay of general principles (e.g., natural selection, physiology, constraints and rules for behavior) with the specific and contingent history of particular populations (Mayr 1959[1976]:317).  The best one can do towards providing a "general" explanation for a given phenomenon is to build increasingly detailed and comprehensive theory and models which show how a set of invariant principles interact with a pool of variability to shape evolution in consistent ways.   Such theories can be *sufficient* for explaining a phenomenon, but never *necessary*, since they cannot specify how the contingent history of a situation interacts with general principles.  In reality, an explanation for a specific instance of cultural elaboration will take the form of a narrative, showing how various general principles interact with the history of a specific population to produce the unique and cumulative history we see in the archaeological record (O'Hara 1988).

       Within the umbrella of Darwinian evolution, several investigators have begun to formulate and champion theories of cultural elaboration (Boone 1998; Dunnell 1989; Neiman 1998).  Our purpose in this chapter is to explore and deepen our understanding of one particular theory, that proposed by Dunnell (1989) in his discussion of "waste" in the archaeological record of eastern North America.  In particular, we attempt to provide a quantitative understanding of Dunnell's "waste" explanation, and to evaluate whether the theory is sufficient for producing populations in which elaboration can be fixed by natural selection.  Evaluating whether the theory is *necessary* and sufficient is outside the scope of the present analysis; as discussed above, such an evaluation requires attempts to apply and falsify hypotheses generated from the theory in the context of a particular evolutionary narrative.  For a beginning to the latter task, we refer readers to the accompanying chapters in the present volume.  After examining Dunnell's (1989) "waste" explanation in some detail, we examine "bet-hedging" models of reproduction in uncertain environments in order to provide a quantitative understanding of his model. The remainder of the chapter details an attempt to assess the completeness and sufficiency of the resulting theory through the use of agent-based simulation modeling.  We conclude by deducing some consequences of the theory for common classes of archaeological evidence, for use in constructing necessary and sufficient explanations of the record.

# EVOLUTION AND CULTURAL ELABORATION

Cultural elaboration is common in many times and places, though certainly not ubiquitous. In an attempt to demonstrate the power of evolutionary theory, Dunnell (1989) examined the distribution of elaboration in eastern North America at the regional scale. Dunnell assumed, as do we, that all organisms use energy in behaviors unrelated to their immediate welfare or reproduction. This assertion seems unremarkable to us, at least for vertebrates. Such expenditures of energy, termed "waste" by Dunnell, are simply present in all individuals at *some* level. That such expenditures are, in fact, unrelated to current survival or reproduction may be controversial to some. Certainly, such expenditures of energy *appear* to provide a conundrum for an adaptationist argument, and probably would remain a paradox if we sought some immediate "function." Instead, the question Dunnell attacked was: are there conditions under which expenditure of energy on elaboration or "waste" could be favored by selection, and thus increase the level of investment within certain populations? The latter question, instead of seeking some immediate function that "wasteful" expenditures of energy serve, simply asks whether there exist selective environments in which "waste" could be a positive contributor to fitness. A very different question indeed.

Dunnell's answer was that at least one set of conditions did exist under which selection could favor "waste" among individuals. If individuals (or families, etc.) vary in the amount of energy they expend in "wasteful" ways, environments with unpredictable variance in resources could favor "wasteful" individuals because these individuals would tend to have lower reproductive rates and thus be less susceptible to resource shortfalls (Figure 1). Dunnell reasoned further that cultural elaboration or "waste" should occur in those regions where resources would be unpredictable, rather than regions where resources simply tended to be either rich or poor. Environmental uncertainty plays a role in evolution because phenotypes take on fitness values only with respect to given environments. When environments vary, fitness values necessarily fluctuate. When fitness values vary predictably, selection can favor suites of traits that "track" the environment, often by facultative response (West-Eberhard 1989). When the environment varies in an unpredictable manner, however, fitnesses may fluctuate unpredictably. Selection typically cannot "track" the environment in an effective fashion under these circumstances. Instead, selection will favor traits which may be sub-optimal in any specific environment, but which yield the highest payoff across the range of environments encountered by an individual (Seger and Brockmann 1987).

Within the Late Archaic and Woodland archaeological record of the Eastern United States, elaboration in the form of burial ceremonialism is frequent though not ubiquitous, either in space or in time. Burial ceremonialism can be traced to two separate lineages in the eastern half of North America. Elaboration in burial appears in the Maritime Archaic of southern Canada (e.g., Tuck 1984). Extensive mortuary activity also occurs separately in the areas south of the Great Lakes as the cultural forms of Glacial Kame and Red Ochre that eventually become what is known as Adena and Hopewell (Caldwell 1958; Cunningham 1948; Dragoo 1963; Griffin 1948; Railey 1990; Ritzenthaler 1957; Webb and Snow 1945). In terms of history, burial ceremonialism first appears along the western and northern margins of the deciduous forest zone, rather than in the biotically richer central or southern oak-hickory forest (Figure 2). This pattern is not what one would expect if elaboration is the product of surpluses or leisure time, as predicted by the cultural evolutionary canon. The pattern is, however, precisely what we would expect given Dunnell's explanation. Elaboration occur earliest in those parts of the eastern forest where mast-producing species occur in low densities compared to forests further east or south, and where resource levels would be the most variable (Buikstra 1981; Chapman 1975; Charles and Buikstra 1983).

Dunnell's "waste" explanation performs well when evaluated against the tasks it set out to accomplish. First, the model accounts for the temporal and spatial pattern of cultural elaboration in prehistoric eastern North America, at least at the regional scales for which data are most readily available. Second, the explanation is a demonstration of how Darwinian analysis could be applied to an important archaeological problem that, at first glance, appears to provide a conundrum for explanation via natural selection. Dunnell's exposition of the "waste" explanation did not, however, specify the exact mechanisms by which the tradeoff he discussed could occur in uncertain environments, nor did he intend to do so in the 1989 paper (Dunnell, personal communication, 1998).

The need for further theoretical development in the current work stems from the desire to test the explanation in the context of specific populations. At this scale, the original model is cast in terms too general to specify the quantitative relationship between reproduction, energy use, and environmental uncertainty for individuals, since Dunnell's focus was on the distribution of phenomena at regional scales. Therefore, the explanation suggests but does not directly refer to the variation in phenotypes within a population that would form the basis for a formal theory of elaboration. The reader should note, however, that we do not consider this a criticism of Dunnell's (1989) discussion at all. Theory consists of a series of explanatory "layers". The upper layers provide the general principles by which the major principles of the theory work, such as natural selection (e.g., Dobzhansky 1970:101-104). Lower layers of theory provide detailed quantitative analyses of how a process works under different sets of conditions (e.g., Gillespie 1991; Roughgarden 1979). Our purpose here is to work within the high level structure of Dunnell's explanation and formulate a theory of the evolution of elaboration suitable for use in drawing hypotheses about individuals and populations.

## NATURAL SELECTION IN FLUCTUATING ENVIRONMENTS

Dunnell's (1989) explanation for elaboration frames the evolution of "waste" as a specific consequence of evolution within uncertain environments. Thus, we believe that the logical source of more detailed theory and quantitative models is the literature concerning natural selection in fluctuating environments, and its extensions in the field of life history theory (Gillespie 1991; Stearns 1992). Life-history theory concerns the explanation of variation in the timing of fertility, growth, and death of organisms (Stearns 1992). A basic tenet is that organisms must always balance the payoffs of behavior with their costs; energy used for one purpose cannot be used for another. There is, for example, often a fitness tradeoff between the number of offspring an individual produces and the probability that they will survive in a variable environment subject to shortfalls. As both theory and empirical research have shown, whenever there is variation in the optimal number of offspring from year to year, fitness is best measured by the geometric mean of reproductive success taken over generations (Gillespie 1977, 1991; Tuljapurkar 1990). Individuals that have more offspring but a higher variance in the survival of those offspring may have lower fitness than individuals who have fewer offspring. This effect, known as bet-hedging, has been observed in many species of animals and plants in unpredictable environments (Boyce and Perrins 1987; Bulmer 1984, 1985; Cohen 1966; Nilsson, et al. 1996).

The bet-hedging idea is based on the notion that in cases for which fitness is a function of environmental condition, heritable phenotypes whose fitness varies less between environmental states may increase in frequency (Seger and Brockmann 1987; Slatkin 1974). This increase is a result of decreased variance in fitness. Moreover, selection in uncertain environments may actually result in the fixation of a phenotypic variant whose *average* fitness is *lower* than that of another trait. This is because the appropriate measure of long-term fitness in temporally variable environments is not the *arithmetic mean* of fitness – its average – but the *geometric mean* of fitness (Gillespie 1977, 1991). When evaluating the success of a phenotypic trait in a population and over time, the geometric mean is the appropriate measure of fitness because reproduction is inherently multiplicative (Gillespie 1991).

A simple example illustrates the effect (Figure 3). Take two strategies, one a high-risk strategy with a high pay-off in reproductive success (shown in Figure 3 as circles) and one a low-risk strategy with a lower pay-off in reproductive success (shown in Figure 3 as squares). During good years in variable environments, the high-risk strategy provides a better payoff than the low-risk strategy. However, in bad years, the pay-off for the high risk strategy is particularly low. Although the arithmetic means of the payoffs are the same, it is the low-risk strategy with a geometric mean of 0.346 that has the higher realized success across any real mix of good and bad years.

Figure 4 demonstrates how the geometric mean effect can impact reproductive strategies such as family size. It has long been recognized that there is an evolutionary trade-off between the number and quality of offspring (e.g., Lack 1947; Cohen 1966; Stearns 1992). The tradeoff occurs when selection favors a reduction in the number of children an organism has at any one time in return for increased probability of survival of those children. Figure 4 depicts the number of children surviving to reproductive age as a function of family size for three different states of a particular environment. In good years, the optimal family size is ten, in intermediate years it is eight, and in bad years it is four. Assuming that the three different environmental states occur with equal frequency, the geometric mean fitnesses of phenotypes that have family sizes of four through ten are represented here by o's, and arithmetic mean fitnesses are represented by x's. The individual with the highest geometric mean fitness in this example is the one with a family size of six. Although this individual is not the one with the highest arithmetic mean fitness and consequently, the highest fecundity or the highest lifetime reproductive output, the dramatically lower overall success of individuals with larger families in bad years results in higher geometric mean fitnesses for individuals which produce smaller family sizes. This exact effect is well documented in other species, with a particularly clear example being Boyce and Perrins (1987) work on clutch size in a well-studied population of birds.

## LIFE HISTORY TRADEOFFS AND THE PHENOTYPIC ENERGY BUDGET

Clearly, the strategy that has the greatest reproductive output is often not the strategy with the maximum fitness. Instead, the most successful strategy may be one with a reduced level of reproductive effort (Sikes 1998). In bad years, then, *any* trait that contributes to producing fewer offspring can ultimately increase fitness. This possibility is very likely given that many cultural behaviors affect the amount of time, energy, or resources an individual may have available for reproduction and parental investment. The real question is whether such expenditures of energy qualify as "bet-hedging." In this chapter, we define "bet-hedging" in its narrow sense as the sacrifice of some fecundity, and thus potential reproductive success, in return for reducing an individual's variance in success as it experiences changing environmental conditions and potentially increasing an individual's cumulative fitness. The remainder of this chapter is aimed at exploring whether, at least in principle, we can construct a formal model in which it is possible for selection in uncertain environments to favor expenditures of energy which do not contribute in an immediate way to reproduction or survival.

Any formal model of behavior we construct must have a well-defined link between reproduction, energy usage, and the effects of the environment upon reproductive success. In order to understand how the "wasteful" life history strategy might be expressed in cultural portions of the phenotype, we must link the bet-hedging model from the previous section with the concept of an organism's "energy budget" (Kooijman 1993). Humans, like all other organisms, operate within an energy budget established by metabolism, costs of reproduction, and energy extraction from the environment. Tradeoffs in the allocation of this energy are inevitable; energy used for one purpose cannot be used for another. In addition, energy investment does not result in ever-increasing returns; the relationship between energy expenditure and returns for most activities is asymptotic – what economists refer to as the law of diminishing returns (Stephens and Krebs 1989).

That reproduction itself impacts the energy budget is incontrovertible; exactly how much and when the "costs of reproduction" are paid by individuals is, however, a matter of great controversy. On a strictly qualitative level, however, in many environments having additional children may stretch the energy budget to the point where the survival of all are threatened. Before this would typically occur, however, selection in variable environments may favor individuals with higher inter-birth intervals, or increased age at marriage or first birth, as discussed earlier. In humans, much of the phenotype for birth spacing and other demographic parameters is at least partially under cultural control and heritable among individuals (e.g., Mace 1998).

The bet-hedging effect may play out as investment in cultural elaboration – that is, in uses of energy that are archaeologically visible as "waste." Burial elaboration, for example, could create a proximate means for the lengthening of the birth interval, since the energy put into preparing the items to be buried with the individual, or creation of a burial structure cannot then be put into parental investment. Determining whether a given expenditure of energy qualifies as a tradeoff is an empirical matter. In order to make this determination, one must assess the effect a behavior or trait has on both average and variance in reproductive success. In addition, in studying how selection operates to favor "wasteful" behaviors we should keep in mind Sober's (1984) concepts of *selection of* and *selection for*. Sober points out that while selection operates upon individuals, creating differential survival and replication, selection operates *as a consequence* of variation in the attributes of individuals. While there is selection *of* individuals, this always occurs via selection *for* properties. In the case of "waste", we would argue that under conditions of marked uncertainty there would be selection *for* traits like increased birth spacing due to increased long-term reproductive success in families with smaller numbers of dependent offspring. Since longer birth spacing might result from investment of energy in activities other than reproduction, such as construction of monumental architecture, or creating burial goods, selection for increased birth spacing would also result in selection *of* individuals with greater investment in "wasteful" behavior.

## ELABORATION AND LIFE HISTORY TRADEOFFS: A FORMAL MODEL

Scientific theories, as discussed above, are often constructed in a series of layers, from general overviews of processes, to families of specific models of processes under different conditions. Our purpose in this chapter has been to explore Dunnell's (1989) "waste" explanation by providing a quantitative basis for his account. A considerable body of theory exists describing the characteristics of natural selection in situations where fitness varies with time as the environment changes; to this theory we have linked Dunnell's explanation through the concept of organismic energy budgets derived from life-history theory and behavioral ecology.

Our goal, however, is not just to point the way towards more specific theory, but to develop the theory to the point where we can derive expectations useful in real archaeological research. The traditional method for doing so in population genetics would be to analyze the properties of a mathematical model of the process in question, deducing expectations for various classes of observations. This process works in genetics, however, because of general agreement about the fundamentals of genetic transmission. There is much less consensus, however, concerning the fundamentals of cultural transmission, despite efforts by Boyd and Richerson (1985) and Cavalli-Sforza and Feldman (1981) to bring the techniques of mathematical modeling to bear on the problem. Fortunately, we have available to us as researchers a different battery of methods -- individual-based simulation techniques allow us to construct a realistic model of individuals and environments and analyze the selection processes that emerge through their interaction (Epstein and Axtell 1996). Though a mathematical theory of bet-hedging in culturally transmitted traits is ultimately needed, it is still possible to derive useful empirical expectations through simulation alone. In the following section, we use simple mathematical models not to represent how we believe selection to act on culturally transmitted traits in a variable environment, but to guide our assumptions and rules when constructing the simulation model.

*A Simple Formal Model of Bet-Hedging in Cultural Transmission*

The hallmark of cultural transmission is the fact that reproduction does not proceed in "generations," but is continuous throughout the lifetime of individuals (Boyd and Richerson 1985; Cavalli-Sforza and Feldman 1981). To model the effects of continuous transmission of traits throughout an organism's life history, Seger and Brockmann's (1987:194) haploid temporal model with overlapping generations and iteroparity is good starting point for deriving some expectations for the behavior of our simulation system. The model presumes discrete but overlapping generations, instead of being a continuous model. Discrete systems have the advantage of being simple to formulate and are usually simple to analyze. Obviously, a discrete model is not a good starting point for a true model of cultural transmission; we argue that the simplification is adequate for deriving important properties of the bet-hedging model that any simulation we construct must have. The formal model presented below is explicitly *not* meant to represent our view of how the geometric mean fitness effect will be incorporated into quantitative theories of cultural transmission. That said, in Seger and Brockmann's model, overlapping generations are created by allowing a proportion (*s*) of the current time period's transmitting individuals to survive and continue to replicate during the next interval. Thus, each time period's population is composed of a relatively large number of individuals that have replicated previously, and a smaller number of "new" individuals (1-*s*). If we consider a population with two traits *A* (*wasteful*) and *B* (*non-wasteful*), with the fitness of *B* scaled to that of *A*, the recurrence equation for each interval is:

**Equation 1**

$$p_{t+1} = p_t [s + (1-s) W_t / V_t]$$
where $W_t$ is the relative fitness of A in generation t, and $V_t$ is the population mean fitness $1 + p_t (W_t - 1)$.

The parameter *s* is also related to the "amount" of environmental variability individuals tend to experience over their lifetimes, since the environment is constantly changing. When *s* is near zero, few individuals survive to transmit in the next time period, and thus experience only a single environmental state. When *s* is near one, however, nearly all individuals survive more than a single time period, thus experiencing more than one environmental state. While hardly faithful to the kinds of complex environments we envision composing the selective environment for elaboration, this method of representing overlapping generations allows us to explore the simplest possible situation in which bet-hedging effects might occur. Assuming a stationary random distribution of environmental states, the equilibrium proportion of A (*wasteful*) is given by the function that maximizes:

**Equation 2**

$$f(P) = s\sqrt{V_1(P)V_2(p)} + (1-s)[P(W_1 W_2)^2 + (1-P)]$$

where P is the expected frequency of $A_1$,
and the $V_i a(P)$ and $W_i$ are the population and
individual fitnesses in the two kinds of periods.

where P is the average frequency of *A*, $V_i(P)$ is the fitness of a population with a given a frequency of A in environmental state *i*, and $W_i$ is the average fitness of individuals of phenotype *A* in environmental state *i*.

When the parameter *s* is near zero, indicating that individuals experience few episodes of environmental variation, the right hand term of Equation 2 contributes the most weight to the proportion of phenotype *A* in the population; that is, what matters is *individual* geometric mean fitness. As *s* increases, the number of fluctuations from "good" to "bad" states increases for each individual throughout their lifetime. When *s* is large, more weight is given to the first term in the equation, which emphasizes the effects of the population's average geometric mean fitness on the frequency of phenotype *A* at equilibrium. Both individual and population geometric mean fitness may favor "wasteful" phenotypes in many environments, at least to some extent. The population's geometric mean fitness, however, may often be maximized where there is a polymorphism of the two strategies, rather than a pure population of either "wasteful" or "non-wasteful" individuals.

Thus, in situations where "generations" often overlap highly (i.e., when the value of *s* can be expected to be high), a shifting balance between wasteful and non-wasteful phenotypes may be expected in the population. Since cultural transmission is a system in which "generations" are relatively continuous, with considerable horizontal and oblique transmission, we would expect that whenever "wasteful" traits are transmitted culturally, we should see a polymorphism of wasteful and non-wasteful phenotypes in the population. In situations where one can measure "waste" as a quantitative variable, in terms of total investment or rates of investment, we would expect to see a smooth, rather than a sharply peaked, distribution. In evaluating whether our simulation is properly constructed to examine tradeoffs caused by culturally transmitted behaviors, we shall look both at the assumptions that go into the model, as well as whether the expectations of Equation 2 are met with respect to polymorphism.

### *Evaluating Theory through Simulation Modeling*

The next necessary step in transforming this general discussion about variability in environments, reproductive behavior and cultural transmission into something more than meta-theory (*sensu* Wasserman 1981) is to expand and explore the sufficiency of the model in detail. In this regard, simulation can be a useful tool for developing specific expectations that investigators can use to develop hypotheses for empirical situations. It is one matter to argue that selection produces changes in the frequencies of a particular attribute; it is quite another to understand how those frequencies dynamically vary in particular environmental and phenotypic conditions. In particular, it is critical that one determines the sufficiency of a complex model before time and money are spent examining its empirical utility. Indeed, collecting primary archaeological data is an extraordinarily expensive and time consuming means for evaluating the ability of a theory to account for classes of phenomena. Simulation is an excellent means for studying the complex interactions of the components of a model, prior to performing expensive and possible destructive analysis.

Simulation modeling, however, has a bad reputation in archaeology, primarily because those using it have not recognized that simulation plays a limited role in science. Simulation is appropriate for deducing the consequences of a theoretical model or set of equations when solving the equations directly is difficult or impossible. One can also use simulation for deriving test implications from a mathematically complex model for testing against empirical data. Simulation modeling should not be used to build "digital replicas" of a prehistoric system or society. Given enough lines of program code, the programmer can replicate virtually any behavior desired in their simulation model – but have we learned anything new by "programming in" all of the behaviors and effects we already knew existed? In our minds, the best use of simulation in science is to explore the complex interactions of a set of *simple* assumptions. For example, if one postulates that natural selection is responsible for the success of a given trait, a simulation that includes an explicit "selection" step will not tell the researcher anything new – selection has been "programmed in" from the very start. In contrast, if we begin with a model where agents inhabit a simple environment, obey simple rules for reproduction, foraging, and other behaviors, it is entirely appropriate to use the simulation to determine when and where natural selection will "emerge" from the interaction of the digital agents. Such a model can be used to determine what environmental and demographic circumstances might create the observed trend in genotypes or cultural replicators, and thus the observed trend in phenotypes. The latter approach, termed "agent-based" or "individual-based" modeling, has been followed by Epstein and Axtell (1996) in their "Sugarscape" model and is the approach we follow here.

Individual-based modeling is a relatively new paradigm in the simulation of systems with many interacting parts (Judson 1994; Langton and Hiebeler n.d). Traditional approaches to simulation tended to represent the behavior of systems of individuals through differential equations representing the modal behavior of individuals taken as a group. Modeling selection in such simulations is unsatisfactory, since one is completely specifying the nature and intensity of selection through the equations. Individual-based modeling offers a different approach, one that allows the dynamics of selection to emerge through the natural interactions between individuals and objects representing their environment. In many ways, a philosophical outgrowth of object-oriented programming methods, individual-based simulation models represent a powerful technique for building and exploring the implications of selection models.

The change in perspective is significant. Researchers have come to appreciate that many of the complex phenomena we see around us are simply the global consequences of local behavior. Such studies have begun to revolutionize many aspects of economics (bringing microeconomics to the fore) and political sciences. In general, individual-based approaches have led to the development of theory that can better account for the stochastic nature of historical change. In anthropology, individual based simulation is just beginning to make inroads. What is clear, however, is that individual based simulations allow anthropology to develop a mathematical formalism that is absent in social sciences and to construct theory that can account for historical and social phenomena. In addition, agent-based simulations permit anthropologists to test their assumptions, such as those posed by evolutionary ecologists, about behavior in complex scenarios and test whether (and under what conditions) these assumptions can generate the classes of phenomena that they predict.

We based our simulations on a programming architecture known as SWARM. SWARM is an emerging standard for agent based modeling that has been under development at the Santa Fe Institute for the past several years. SWARM is unique in that it permits scientists to create very complicated models and to explore aspects of multi-dimensional interaction (requiring sophisticated programming) with minimal effort on the programmatic "mechanics" of the actual application (e.g., memory management, display management, etc.) For this reason, SWARM is serving as a central focus to a wide range of researchers in fields such as anthropology, economics, biology, physics and archaeology. A number of archaeologists have begun using SWARM to examine issues such as the formation of villages and the effects of environment on agriculture (e.g., Kohler and Carr 1996).

*Elements of a Simulation Model for Waste as Reproductive Tradeoff*

To model life history tradeoff predictions in a variety of environmental conditions and to test whether (and under what conditions) life history models can actually generate the classes of "wasteful" phenomena that they predict, we built an agent based simulation of these processes. The basic components of our waste simulation consist of a population of agents with variable phenotypes and a variable environment that consists of a single resource, arbitrarily called sugar. Agents are given the ability to move, forage for food to meet metabolic needs and a set of rules for interacting with others and reproducing new agents throughout their lifetimes. The features of the simulation include biological reproduction, realistically uncertain environments, and phenotypes composed of "genetically" and "culturally" transmitted traits.

To examine the effects of unpredictable environments on reproductive success, the rules of reproduction are an important component of our model. When agents reach a model-specified minimum age for reproduction, they can reproduce provided they meet several biologically and culturally determined conditions. The conditions under which reproduction is possible are sex specific. In order to reproduce, both females and males must possess a biologically determined minimum amount of energy as well as an additional amount that is determined by a culturally inherited preference. Females also have a biologically and culturally determined amount of time they must wait between births (i.e., birth spacing). If an agent that is ready to reproduce meets an agent of the opposite sex who is also ready to reproduce, a new agent is born. This agent inherits biological parameters from its parent in a simple Mendelian manner (without crossover), and inherits an initial random sample of its parent's cultural repertoire. In order to assign a cost to having children, the simulation required parents to be responsible for providing children sugar resources until they are old enough to forage on their own.

A second key feature to our simulation is cultural transmission. The phenotypes of agents were modeled to be composed of traits that were transmitted both culturally and genetically and were generated as follows. As agents move around the landscape, they randomly encounter one another. When encounters occur, there is a probability that the agents will "talk" to each other and exchange cultural traits. In the waste simulation, cultural traits are modeled as "tokens" that can be one of three types. The first type can be taken without cost to an agent. The second type invokes a cost in sugar to the receiving agent. Agents are not required to take these tokens but are given a cultural rule that determines the maximum token cost that an agent is willing to pay. The third kind of tokens are the ones which code for cultural preferences for time between births, the amount of sugar required before having a child, and the maximum token cost the agent is willing to pay. These tokens have no cost, and result in the replacement of the receiver's preference by the preference of the transmitter.

Because tokens flow culturally and genetically through the population independent of one another, persistent phenotypes are emergent properties of token combinations that individuals possess at any given moment in time. To account for the effect of agents with constantly changing phenotypes, the simulation tracks the reproductive success of phenotypes, rather than individual agents. Change in the distribution of phenotypes is the result of differential persistence and transmission of tokens due to selection and drift and *not* a function of decision-making rules embedded in the agents.

We have defined "wasteful" traits as behavior or structures that have a cost in the short run but a benefit in reproductive success over the long run by lowering the variance of fitness. In our simulation, we examined the distribution of values across three variables of interest: the agent's inter-birth interval, the amount of sugar stored and the energy spent on cultural tokens.

Phenotypes for tracking the amount of "wasteful" behavior were created by treating each variable as a dimension, dividing each dimension into a series of modes and by creating a paradigmatic classification. Though selection is not explicitly programmed into the simulation, our model predicts that the frequencies of phenotypes in the population will change as a consequence of differential reproduction and cultural transmission. In order to examine the effect of selection on the frequency of these phenotypes, agents were subjected to a suite of environmental conditions in which the rate of sugar growth was varied. The kinds of environments we studied included constant growth, cyclical, chaotic, and environments with periodic failures. The effect of spatial variability and mobility was examined by allowing agents to move greater or lesser distances to search for resources. Tracking the number of children each individual produced and their lifetime expression of "wasteful" behaviors permitted us to calculate the geometric and arithmetic mean fitnesses for phenotypes. In addition, we tracked changing patterns of age structure of the population, population size and variances over time, as well as the distribution of wasteful behaviors in the population.

## RESULTS

Though the Swarm simulation we constructed contains relatively few dimensions along which individuals can vary, the parameter space of "possible" simulation runs is still enormous. Coupled with the fact that we were examining selection within randomly varying environments, the vastness of the parameter space means that the results discussed below are necessarily preliminary, even though they are the results of observing large numbers of simulation runs over a wide set of parameters sampled from the available space. Nevertheless, the simulation runs shed light on the relationship between "wasteful" behavior and environmental effects, mobility of agents, age structure, and the distribution of wasteful phenotypes across populations.

*The Effect of Environmental Uncertainty*

The most general result of the model is that marked unpredictability in the environment is indeed capable of creating selection for "wasteful" behavior within the simulation populations (Figure 5). This slide shows the results from four different simulations in two different environments – predictable (left) and unpredictable (right) with two different populations. One population was composed only wasteful phenotypes while the other population was made up of only non-wasteful phenotypes. In a constant environment, both the arithmetic and geometric mean fitnesses of non-wasteful phenotypes are higher than those of the wasteful phenotypes. In unpredictable environments, however, "wasteful" phenotypes have higher geometric mean fitnesses than non-"wasteful" variants, all other things being equal. Thus, as the mathematics of the bet-hedging hypothesis predict, the general premise of the "waste" model appears to hold true. Additionally, we believe that the model is correctly constituted, since rarely were populations driven to fix either "wasteful" or non-wasteful phenotypes; under all reasonable circumstances the population was composed of a mixture of different levels of investment in "waste."

*Mobility*

Unlike purely mathematical formulations of bet-hedging, however, agent-based simulation allows one to examine more complicated implications of the model such as migration. Migration tends to have an ameliorating effect on the tradeoff between the total number of children and the number of surviving children. That is, individuals can lessen the effects of uncertainty by moving from an area of low productivity to one of high productivity. This slide shows summarizes a set of runs designed to examine the effects of mobility. The simulation runs demonstrate that populations of agents that are given the ability to see and move over larger distances, evolve lower levels of "waste" (Figure 6, on the right of the graph) than populations that are more restricted in their movement. This finding is consistent with what evolutionary biologists have observed with respect to the bet-hedging effect in other species. It also potentially informs on the relationship between cultural elaboration and sedentariness. It has often been argued that cultural elaboration requires surplus generated by resource intensification and permanent settlement. In the simulation, however, levels of wasteful behavior became fixed within the population despite the fact that none of the agents were immobile and dependent upon a single location in the environment for subsistence. This effect demonstrates that sedentariness, as it is usually thought of, is not required for selection to favor "wasteful" phenotypes. Sedentariness merely increases the strength of selection for waste in unpredictable environments.

*Demographic Structure*

Selection for "wasteful" phenotypes was also linked to the average age distribution of the simulated populations (Figure 7). Populations with higher frequencies of "wasteful" behavior tend to have more individuals surviving to older ages through decreased mortality related to the lowered "costs of reproduction." Thus, the tradeoff effect acts not only to increase the geometric mean fitness of the population, but also to alter the age-distribution of the population through increased survival due to decreased reproductive effort. This outcome is illustrated in this slide by the right-skewed age distribution in populations with a greater proportion of "wasteful" phenotypes.

*Phenotypic Polymorphism*

Finally, we found that could measure the degree to which a "wasteful" phenotype can co-exist with other phenotypes in a stable polymorphism (Figure 8). To do this, we scaled phenotypic dimensions such as culturally transmitted birth spacing intervals, accumulation of sugar, and average expenditure of energy on "expensive" culturally transmitted traits along an index of "wastefulness." Placing the populations into predictable and unpredictable environments, we examined the population distribution across this index. As the slide demonstrates, while the *mean* value of this "wastefulness" index is not affected by selection, markedly unpredictable environments yield phenotypic distributions that are strongly *right-skewed*, or skewed in the direction of having more "wasteful" phenotypes. This finding has significant potential for archaeologists seeking to measure degrees of "wastefulness" in the archaeological record. Given the difficulty of measuring the "wasteful" behavior of individuals in archaeological circumstances, these results show that the overall shape of a distribution may be a more appropriate and facile measure than the mean.

### *Consequences for Empirical Research*

Although selection can favor costly artifact classes in variable environments due to the bet-hedging effect, the model does not specify the form that such artifacts can take. "Wasteful" artifact classes will follow historically contingent trajectories within each cultural tradition. Determining the form that any particular instance of "waste" takes is a matter of historical analysis that requires examining a particular dataset in a particular ecological setting.

Additionally, the form that "wasteful" artifacts take potentially provides the variability for other kinds of selective processes. For example, artifacts involved in life-history tradeoffs may also be related to costly signaling, functional specialization and redistribution. It is important to recognize that because of diminishing returns for any one kind of energy expenditure, there are often multiple evolutionary solutions for reducing variance and creating the life history tradeoff effect. The fixation of any particular trait may require additional fitness consequences resulting from food redistribution and other kinds of functional organizations.

In such cases, these proximate mechanisms can act to intensify selection for costly artifact classes. Increased investment in mound building, for example, may be driven by the bet-hedging effect. However, the fixation of mound building within the population may be due to its role in creating a large-scale food sharing system. An important lesson that archaeologists can learn from these results is that they should pay close attention to the frequency of wasteful traits *within* populations, which may be the result of a suite of varied and complex evolutionary forces.

In addition to examining the historical trajectories for artifact classes, archaeologists must also be aware of the role that subsistence systems play in creating the selective environment for "wasteful" traits. The ability to store resources and buffer shortfalls with multiple sources of food, for example, potentially decreases the bet-hedging effect and thus selection for "wasteful" traits. On the other hand, populations that depend upon a single food staple or live sedentary, dispersed settlements may provide particularly strong selective environments for "wasteful" traits since small changes in productivity have large impacts on these kinds of settlement systems. Also, as we mentioned earlier, one clear result of the model is that "wasteful" traits will increase as mobility decreases within an uncertain environment due to geographic restrictions, population in-filling or changes in settlement patterns. Thus, there should be a clear relationship between the level of "wastefulness" seen in populations and the details of settlement strategies.

Like the study of subsistence systems, skeletal data may also play a role in examining how the bet-hedging effect is expressed within a particular population. As has been discussed, life history tradeoffs manifest themselves in biological variables, such as population age distribution and birth spacing intervals (e.g., Katzenberg, 1996; Skinner, 1997). Some of these variables should be measurable in carefully controlled skeletal samples through estimates of age at first birth and age at death, though considerable additional simulation work needs to be done to focus on how the effects discussed above would be manifested, if at all, through the taphonomic and analytic filters imposed by most skeletal death assemblages.

Finally, fine-grained studies of environmental conditions will provide one of the strongest avenues of information for studying the bet-hedging effect in the archaeological record. New high-resolution studies of past climatic conditions provided by ice cores, deep sea drilling, tree rings and coral growth records will undoubtedly serve as a rich source of data for archaeologists examining the selective conditions for "wasteful" phenotypes (e.g., Meeker et al. 1997; Melice and Roucou 1998). These new, yearly and decadal level records of past rainfall and temperature can easily generate information about the amplitude, frequency and magnitude of environmental failures. It is important to recognize that single or isolated environmental downturns are not sufficient to create the bet-hedging effect. In order for there to be selective pressure for wasteful behavior, individuals must experience at least several environmental perturbations. It is the *transition* between good and bad periods of environmental productivity, and thus fitness, that creates variance in success for wasteful vs. non-wasteful phenotypes. Studies of environmental variability should focus on the examination of the variability using such statistics as the coefficient of variance, which consider the effect of changes in means on the absolute amount of variance. In addition, there cannot be an absolute requirement for the minimum amount of variability that populations must experience for there to be selection for wasteful 'phenotypes'. The amount of variability necessary to generate the bet-hedging effect is always a product of population densities, subsistence systems, settlement strategies and the overall productivity of the environment.

## CONCLUSIONS

In this chapter, we have attempted to use a simple mathematical model and agent-based simulation to understand the conditions under which selection would favor the life history tradeoff that may be expressed in "wasteful" phenotypes. It is important to recognize that the explanation provided here is *sufficient* but not *necessary* for any particular case of cultural elaboration. First, given any particular case, there are many evolutionary solutions to coping with unpredictable environments only some of that result in the expression of wasteful behaviors. No individual must engage in wasteful behaviors in order to cope evolutionarily with variability in the environment. Second, cultural elaboration may be favored by selection for reasons other than unpredictable environments. These selective environments potentially include selection for costly signaling, functional specialization and functional integration. Understanding if the bet hedging effect is responsible for increasing investment in wasteful behavior is an empirical matter that can be solved by looking at empirical effects deduced from the model, such as skeletal age distributions.

That explanations of cultural phenomena are sufficient but not necessary also means that it will never be possible to examine the characteristics of a given environment and predict the equilibrium frequency of "wasteful" phenotypes. The form of "wasteful" behavior or artifacts is entirely historical contingent. Rather than being predictive, the life history tradeoff hypothesis is a relatively simple *null* model for the expected distribution of traits related to the bet-hedging effect in the archaeological record. Increasing the sophistication of the simulation may well enhance the model's ability to account for variability in cultural elaboration in space and time, however. For example, although there may be complicated rules for translating inherited information into behavior (in other words, decision making algorithms), this initial model was purposely built to be very simple. That is, we were seeking to determine if the actions of transmission and selection are adequate for generating the conditions necessary to favor "wasteful" kinds of phenotypes. More sophisticated equations for translating phenotypic variables into behavior along the lines of those built by Boyd and Richersen (1985) Cavalli-Sforza and Feldman (1981) or even the newer work on "meme" theory by Gabora (1996) and others (e.g., Lynch 1996; Lynch and Baker 1986, 1993, 1994; Payne 1996; Pocklington and Best 1997) may increase the sufficiency of the simulation. In future work, we will examine the role of the life-history tradeoff model in generating the variability necessary for the evolution of costly signaling, functional specialization and redistribution systems.

**ACKNOWLEDGEMENTS**

First and foremost, we gratefully acknowledge the inspiration, comments, and corrections given us by R.C. Dunnell.  Obviously, this work is an outgrowth of his earlier work, so in a very real sense this chapter would not exist without his help.  We are also thankful for comments on the manuscript by Deborah Schechter as well as discussion by Eric A. Smith and his students.  Portions of the original research for this chapter were performed by Madsen with support from the National Science Foundation Graduate Fellowship program, and with assistance from Sigma Xi.

# REFERENCES CITED

Boone, James
    1998    The evolution of magnanimity - When is it better to give than to receive*? Human Nature* **9:**1-21.

Boyce, M.S., and C.M. Perrins
    1987    Optimizing Great Tit Clutch Size in a Fluctuating Environment. *Ecology* **68:**142-153.

Boyd, R., and P.J. Richerson
1985    *Culture and the Evolutionary Process*. University of Chicago Press, Chicago.

Buikstra, J.E.
1981    Mortuary Practices, Paleodemography, and Paleopathology: A Case Study from the Koster Site (Illinois). In *The Archaeology of Death*, edited by R. Chapman, I. Kinnes, and K. Randsborg, pp. 123-132. Cambridge University Press, Cambridge, England.

Bulmer, M.G.
1984    Delayed Germination of Seeds: Cohen's Model Revisited. *Theoretical Population Biology* **26:**367-377.

Bulmer, M.G.
1985    Selection for Iteroparity in a Variable Environment. *American Naturalist* **126:**63-71.

Caldwell, J.R.
1958    *Trend and Tradition in the Prehistory of the Eastern United States*. American Anthropological Association Memoir.

Cavalli-Sforza, L. L., and M.W. Feldman
1981    *Cultural Transmission and Evolution: A Quantitative Approach*. Monographs in Population Biology No. 16. Princeton University Press, Princeton, N.J.

Chapman, C.H.
1975    *The Archaeology of Missouri I*. University of Missouri Press, Columbia.

Charles, D.K., and J.E. Buikstra
1983    Archaic Mortuary Sites in the Central Mississippi Drainage: Distribution, Structure, and Behavioral Implications. In *Archaic Hunters and Gatherers in the American Midwest*, edited by J.L. Phillips and J.A. Brown, pp. 117-145. Academic Press, New York.

Cohen, D.
1966    Optimizing Reproduction in a Randomly Varying Environment. *Journal of Theoretical Biology* **12:**110-129.

Cunningham, W.M.
1948    *A Study of the Glacial Kame Culture in Michigan, Ohio, and Indiana*.

Dobzhansky, T.

1970     *Genetics of the Evolutionary Process*. Columbia University Press, New York.

Dragoo, D.W.
1963     *Mounds For The Dead*. Carnegie Museum, Pittsburgh.

Dunnell, R. C.
1989     Aspects of the Application of Evolutionary Theory in Archaeology. In *Archaeological Thought in America*, edited by C. C. Lamberg-Karlovsky, pp. 35-99. Cambridge University Press, Cambridge.
1980     Evolutionary theory and archaeology. *Advances in Archaeological Method and Theory* **3:**35-99.

Gabora, L.
1997     A Day in the Life of a Meme. In *The Nature, Representation, and Evolution of Concepts*, edited by Philip van Loocke. Routledge Press, New York.

Gillespie, John
1977     Natural Selection for Variances in Offspring Numbers: A New Evolutionary Principle. *American Naturalist* **111:**1010-1014.
1991     *The Causes of Molecular Evolution*. Oxford University Press, New York.
1998     *Population Genetics: A Concise Guide*. John Hopkins University Press, Baltimore.

Judson, O. P.
1994     The Rise of Individual-based Models in Ecology. *TREE* **9:**14.

Katzenberg, M. A., D. A. Herring, and S. R. Saunders
1996     Weaning and infant mortality: Evaluating the skeletal evidence. *Yearbook of Physical Anthropology, Yearbook Series Vol 39* **39:**177-199.

Kohler, T.A. , and E. Carr
1996     *Swarm-based Modelling of Prehistoric Settlement Systems in Southwestern North America*. Paper presented at Paper presented at the Archaeological Applications of GIS Colloquium II, Sept. 1996, Forli Italy.

Koojiman, S.A.L.M.
1993     *Dynamic energy budgets in biological systems : theory and applications in ecotoxicology*. Cambridge University Press, Cambridge.

Lack, D.
1947     *Darwin's Finches*. Cambridge University Press, Cambridge.

Langton, C. , and D. Hiebeler
n.d.     SWARM Simulation Platform for the Simulation of Complex Systems. Santa Fe Institute, Santa Fe.

Lynch, A.
1996     The Population Memetics of Birdsong. In *Ecology and Evolution of Acoustic Communication in Birds*, edited by Donald E. Kroodsma and Edward H. Miller, pp. 181-197. Cornell University Press, Ithaca, New York.

Lynch, A. and A. J. Baker
1986    Congruence of morphometric and cultural evolution in Atlantic island chaffinch populations. *Canadian Journal of Zoology* **64:**1576-1580.
1993    A population memetics approach to cultural evolution in chaffinch song: meme diversity within populations. *American Naturalist* **141:**597-620.
1994    A population memetics approach to cultural evolution in chaffinch song: differentiation among populations. *Evolution* **48:**351-359.

Mace, R.
1998    The coevolution of human fertility and wealth inheritance strategies. *Philosophical Transactions of the Royal Society of London Series B-Biological Sciences* **353:**389-397.

Madsen, M. E.
in press Evolutionary Bet-Hedging and the Hopewellian Cultural Climax. In *Posing Questions for a Scientific Archaeology*, edited by Terry L. Hunt, Carl P. Lipo, and Sarah L. Sterling. University of Utah Press, Salt Lake City.

Mayr, E.
1959    Where are we? *Cold Spring Harbor Symposium on Quantitative Biology* **24:**409-440.

Meeker, L. D., P. A. Mayewski, M. S. Twickler, S. I. Whitlow, and D. Meese
1997    A 110,000-year history of change in continental biogenic emissions and related atmospheric circulation inferred from the Greenland Ice Sheet Project Ice Core. *Journal of Geophysical Research-Oceans* **102:**26489-26504.

Melice, J. L., and P. Roucou
1998    Decadal time scale variability recorded in the Quelccaya summit ice core delta O-18 isotopic ratio series and its relation with the sea surface temperature. *Climate Dynamics* **14:**117-132.

Neiman, F.
1997    Conspicuous Consumption as Wasteful Advertising: a Darwinian Perspective on Spatial Patterns in Classic Maya Terminal Monument Dates. In *Rediscovering Darwin: Evolutionary Theory in Archaeological Explanation*, edited by C. Michael Barton and G.A. Clark, pp. 267-290. American Anthropological Association, Arlington, VA.

Nilsson, P., J. Tuomi, and M. Astrom
1996    Bud Dormancy as a Bet Hedging Strategy. *American Naturalist* **147:**269-281.

O'Brien, Michael J., and Thomas D. Holland
1990    Variation, Selection and the Archaeological Record. *Archaeological Method and Theory* **2:**31-80.

O'Hara, R.J.
1988    Homage to Clio, or, toward an historical philosophy for evolutionary biology. *Systematic Zoology* **37:**142-155.

Payne, Robert B.

1996    Song traditions in Indigo Buntings: Origin, Improvisation, Dispersal, and Extinction in Cultural Evolution. In *Ecology and Evolution of Acoustic Communication in Birds*, edited by Donald E. Kroodsma and Edward H. Miller, pp. 198-220. Cornell University Press, Ithaca, New York.

Pocklington, R., and M. L. Best
1997    Cultural evolution and units of selection in replicating text. *Journal of Theoretical Biology* **188:**79-87.

Railey, J.A.
1990    Woodland Period. In *The Archaeology of Kentucky: Past Accomplishments and Future Directions*, edited by D. Pollack, pp. 247-374. Kentucky Heritage Council State Historic Preservation Comprehensive Plan Report No. 1.

Ritzenthaler, R.E.
1957    The Old Copper Culture of Wisconsin. *Wisconsin Archaeologist* **38:**183-332.

Roughgarden, Jonathan.
1979    *Theory of population genetics and evolutionary ecology : an introduction*. Macmillan, New York.

Seger, J., and H.J. Brockmann
1987    What is Bet-Hedging? In *Oxford Surveys in Evolutionary Biology*, edited by P.H. Harvey and L. Partridge, pp. 182-211. Oxford University Press, Oxford.

Sikes, R.S.
1998    Unit pricing: Economics and the evolution of litter size. *Evolutionary ecology* **12:**179.

Skinner, M.
1997    Dental wear in immature Late Pleistocene European hominines. *Journal of Archaeological Science* **24:**677-700.

Slatkin, M.
1974    Hedging One's Evolutionary Bets. *Nature* **250:**704-705.

Sober, E.
1984    *The Nature of Selection: Evolutionary Theory in Philosophical Focus*. MIT Press, Cambridge, MA.

Stearns, Stephen C.
1992    *The evolution of life histories*. Oxford University Press, Oxford.

Stephens, David W., and John R. Krebs
1986    *Foraging Theory*. Princeton University Press, Princeton, N.J.

Tuck, J.A.
1984    *Maritime Provinces Prehistory*. National Museum of Canada, Ottawa.

Tuljapurkar, Shripad
1990    *Population dynamics in variable environments*. Springer Verlag, New York.

Wasserman, Gerhard D.
1981   On the Nature of the Theory of Evolution. *Philosophy of Science* **48:**416-437.

Webb, W.S., and C.E. Snow
1945   *The Adena People*. University of Kentucky, Lexington.

West-Eberhard, M.J.
1989   Phenotypic Plasticity and the Origins of Diversity. *Annual Review of Ecology and Systematics* **20:**249-278.

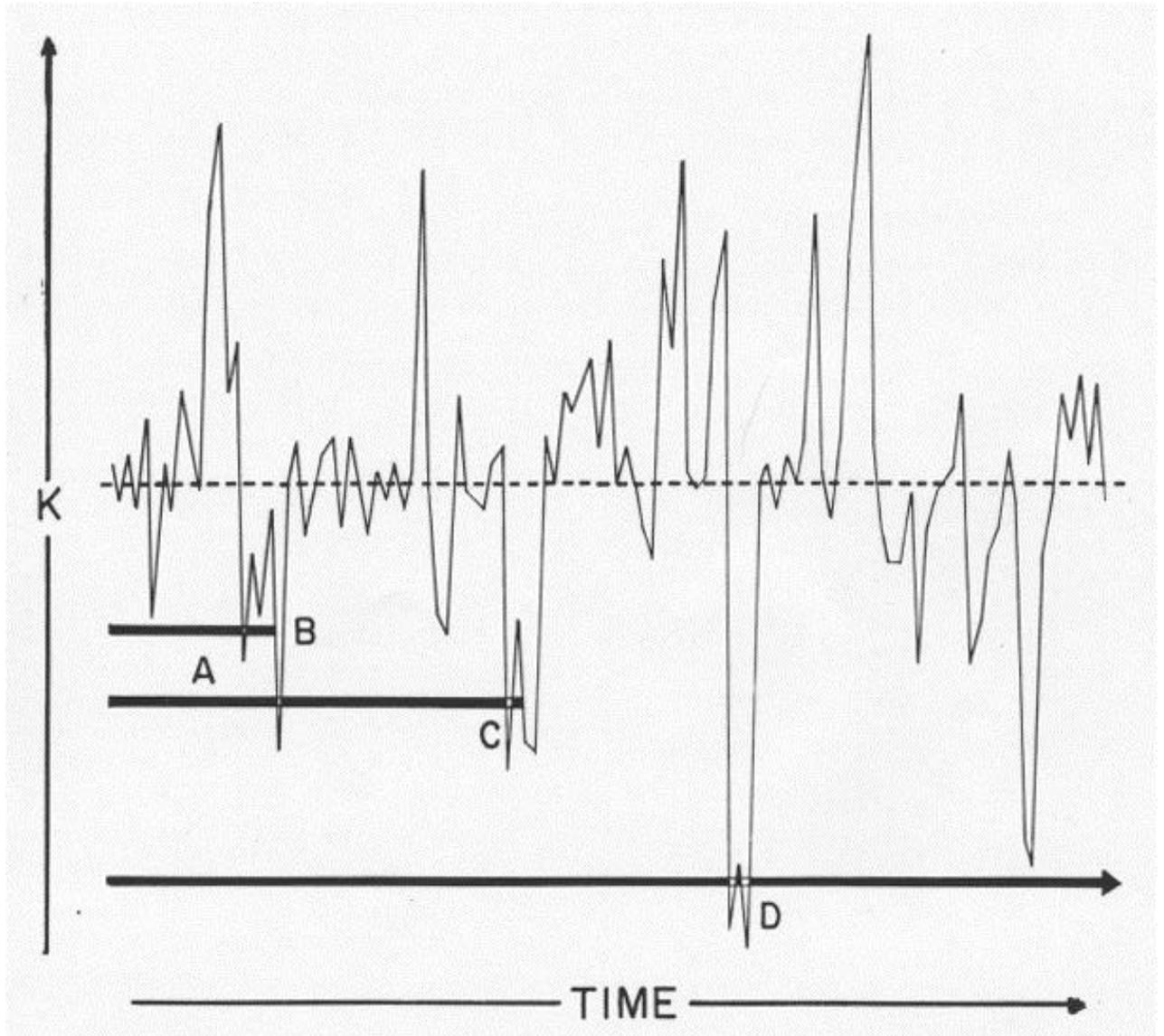

Figure 1: Simplified model of the selection of "waste" from Dunnell (1989). Dunnell conceptualized waste as a behavior not involved in reproduction and that can act as an energy buffer in times of environmental shortfall. Here, mean "carrying capacity" (----) is held constant (i.e. no environmental trends and no change or difference in subsistence) while the carrying capacity at any point in time is highly variable. Three horizontal bars represent different populations in equilibrium at different sizes. All populations can persist through minor shortfalls in productivity (A); however, drastic shortfalls (B) and/or repetitive shortfalls (C) will cause extinction or emigration. In the short run, larger populations are more fit but, in environments which experience large unpredictable fluctuations in "carrying capacity," populations stabilized at smaller sizes by waste-type behavior will be at an advantage not only because of the smaller size but also because temporary abandonment of waste provides a reservoir of time to allow "intensification" (D). Although Dunnell's model appears to account for more of the archaeological record than leisure time hypotheses, questions have been raised about its apparent requirements for group selection.

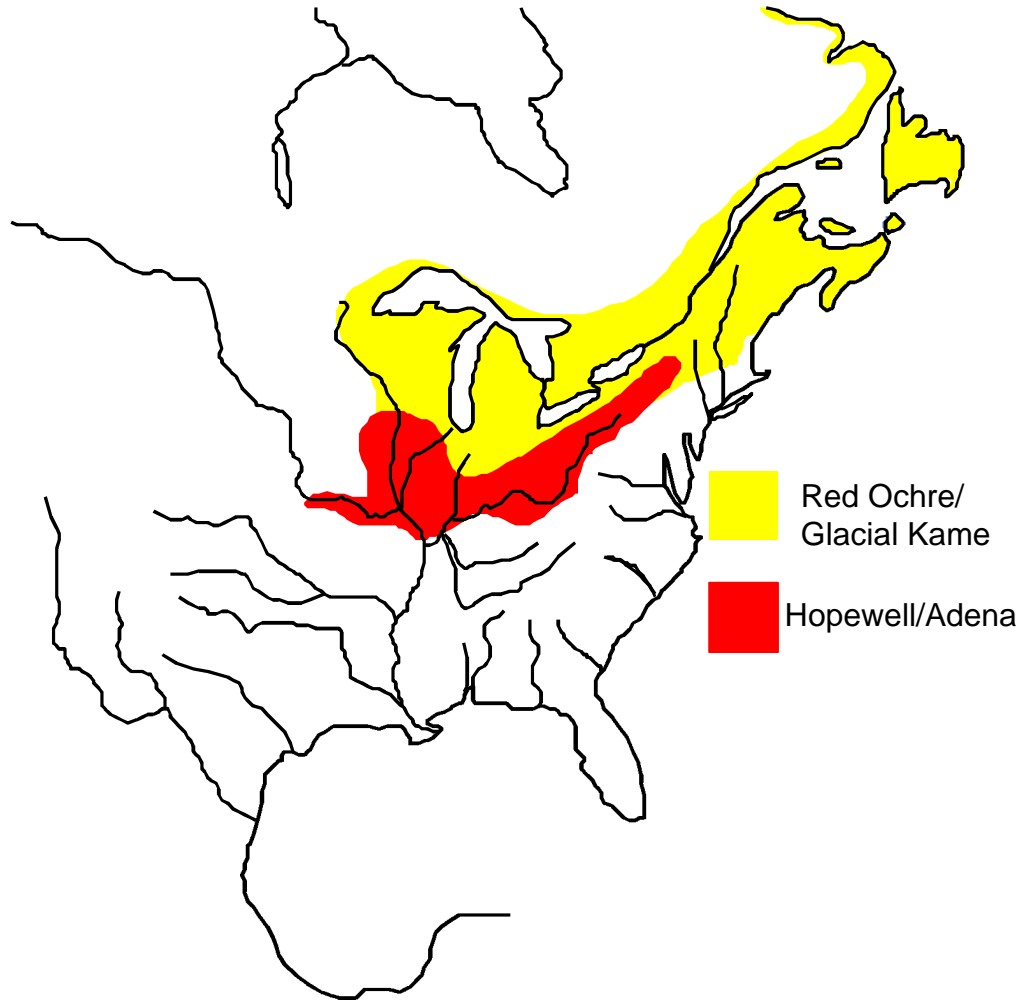

Figure 2: The distribution of Hopewell/Adena and the Red Ochre/Glacial Kame mortuary complexes corresponds closely to the northern, marginal and energetically unpredictable edge of the oak-hickory forests with the earliest evidence occurring in the most northerly and marginal environments.

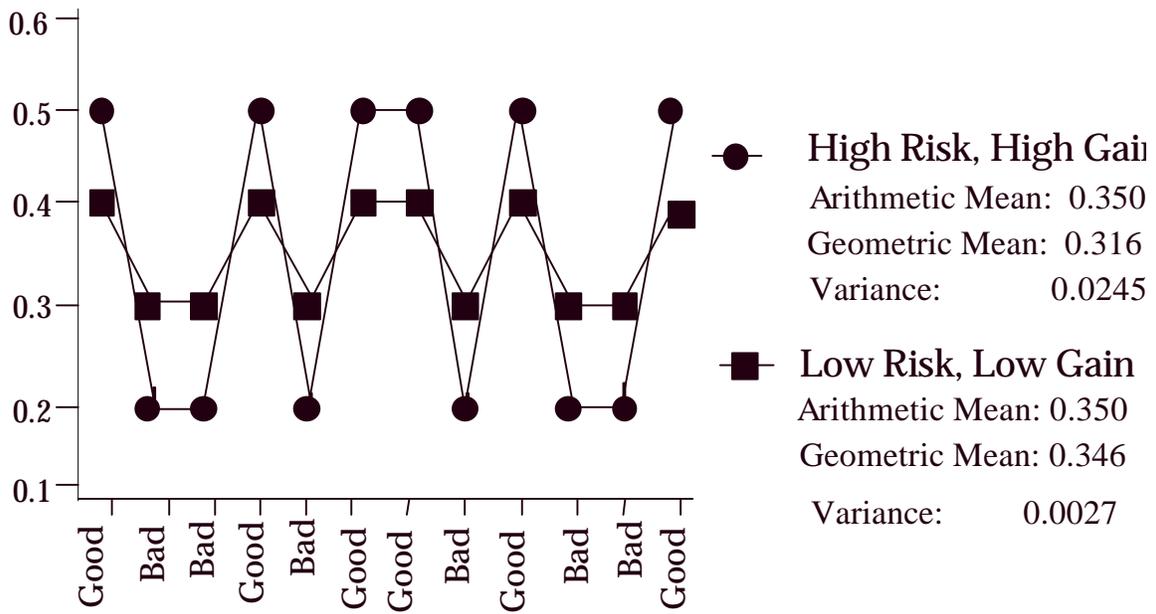

Figure 3: Bet-hedging: Geometric and Arithmetic Means

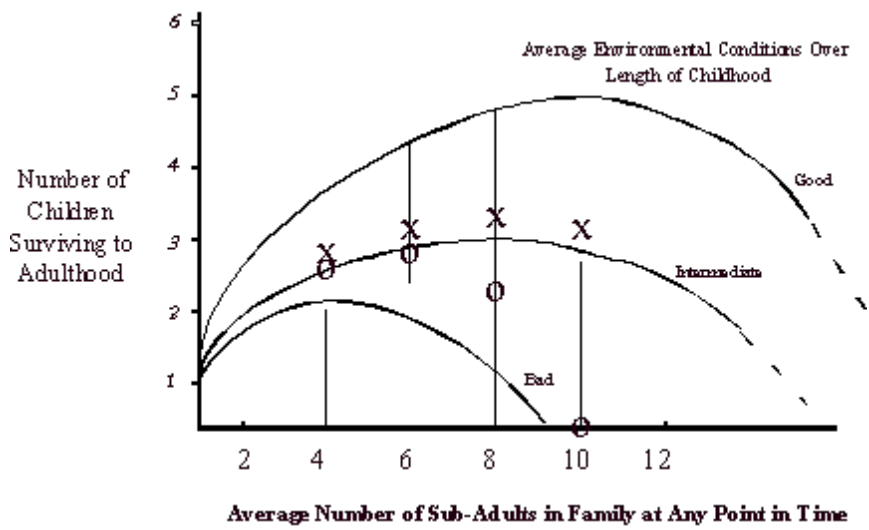

Figure 4: The "Bad Years" Effect on Reproductive Behavior (from Boyce and Perrins 1987).

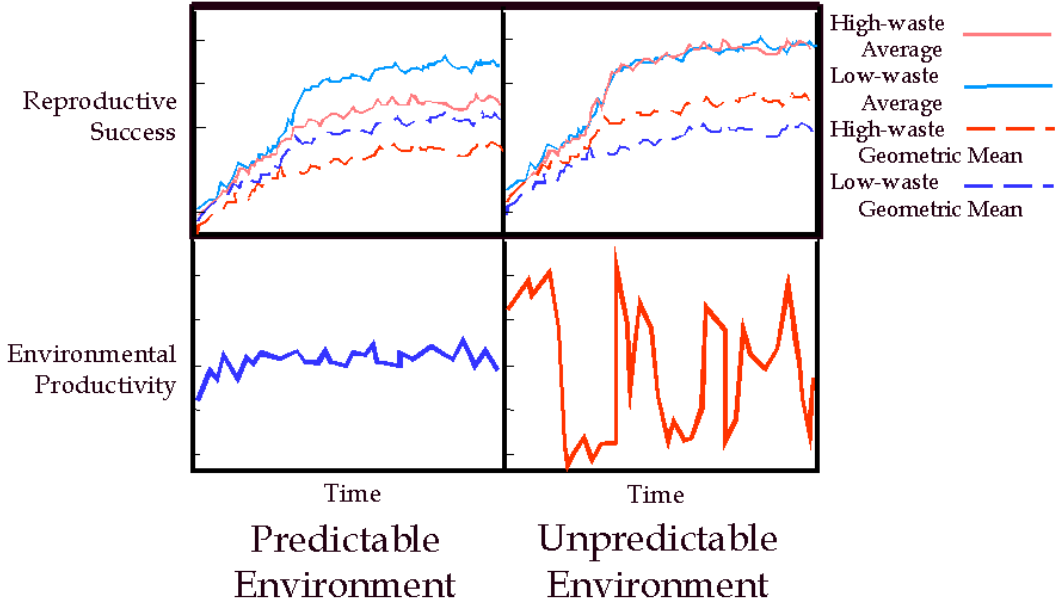

Figure 5: "Wasteful" phenotypes are more reproductively successful in variable environments than less-"wasteful" phenotypes.

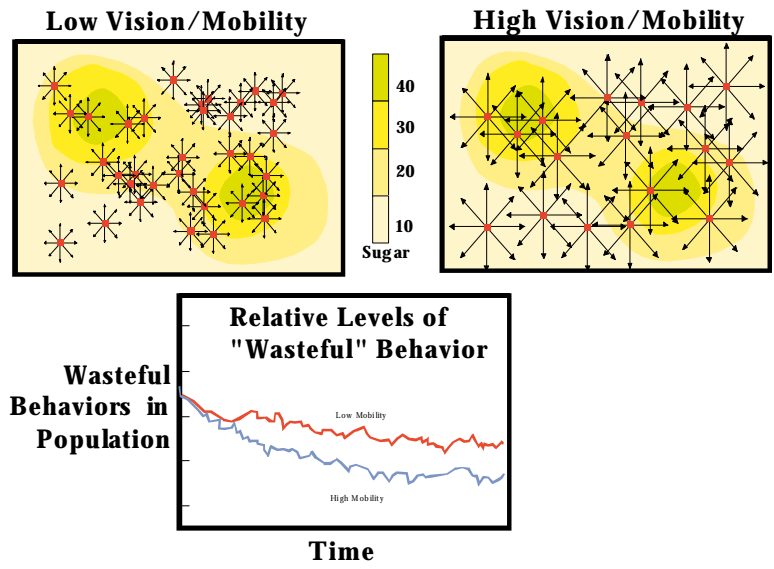

Figure 6: The effect of mobility on selection for "wasteful" behavior.

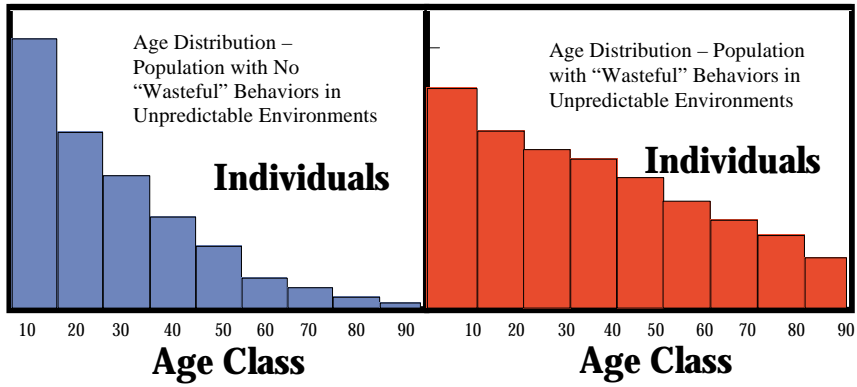

Figure 7: The Effect of "Wasteful" Behaviors on Population Mortality Profiles

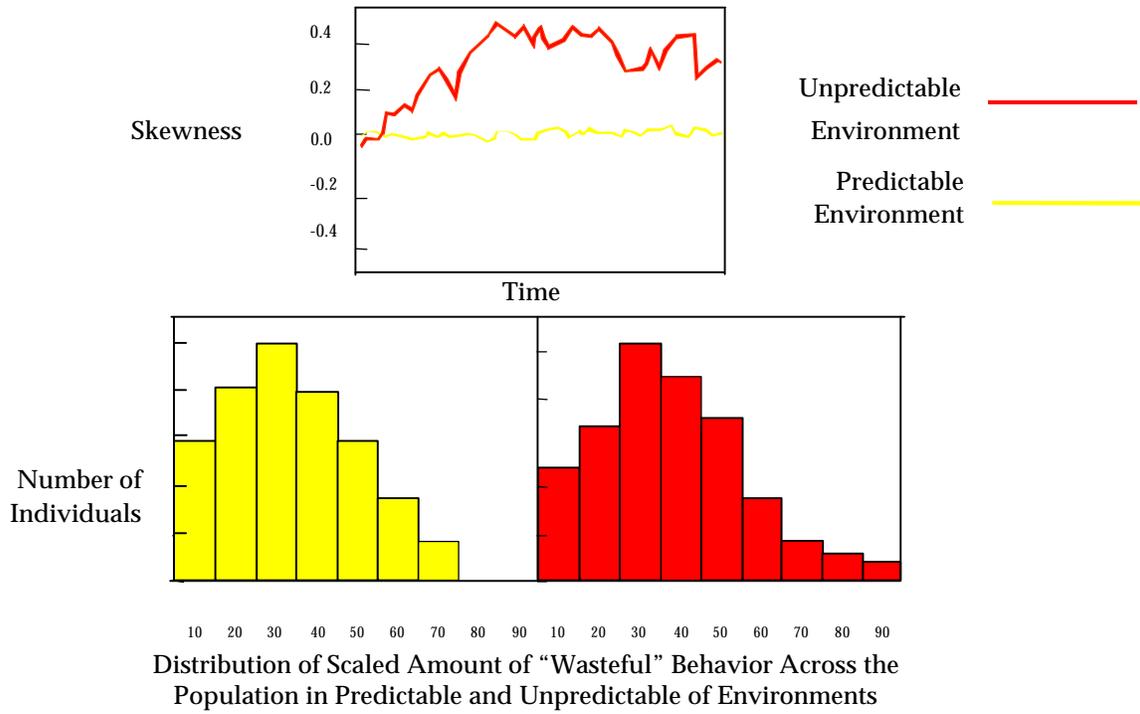

Figure 8: Polymorphic population compositions in predictable and unpredictable environments.